\documentclass[twocolumn,prd]{revtex4-2}
\usepackage{graphicx}
\usepackage{dcolumn}
\usepackage{multirow}
\usepackage{amsmath,amsthm,amssymb}
\usepackage{hyperref}

\def\pdt{\partial_t}
\def\pdr{\partial_r}
\def\pdu{\partial_u}
\def\pdtt{\partial_{t,t}}

\def\pdrr{\partial_{r,r}}
\def\pdru{\partial_{r,u}}
\def\pduu{\partial_{u,u}}
\def\kru{K^r_{\,\,\,u}}
\def\krr{K^r_{\,\,\,r}}
\def\kuu{K^u_{\,\,\,u}}
\def\kff{K^\varphi_{\,\,\,\varphi}}
\def\br{\beta^r}
\def\bu{\beta^u}

\def\tildeta{\tilde{\eta}}
\def\tildkrr{\tilde{K}^r_{\,\,\,r}}
\def\tildkru{\tilde{K}^r_{\,\,\,u}}

\def\tildll{\tilde{\lambda}}
\def\tildbr{\tilde{\beta}^r}
\def\tildbu{\tilde{\beta}^u}
\def\psinum{\left(\psi-1\right)}
\def\chinum{\left(\chi-1\right)}
\def\hl{\tilde{h}^{(\ell)}}
\def\Phil{\Phi_{\ell}}
\def\xil{\tilde{\xi}^{(\ell)}}
\def\hltwo{\tilde{h}^{(\ell=2)}}
\def\Philtwo{\Phi_2}
\def\xiltwo{\tilde{\xi}^{(\ell=2)}}
\def\hlzero{\tilde{h}^{(\ell=0)}}

\def\xilzero{\tilde{\xi}^{(\ell=0)}}
\def\cltwo{c^{(\ell=2)}}
\def\clzero{c^{(\ell=0)}}
\def\mp{M_{lin}}
\def\madm{M_{ADM}}
\def\ub{\bar{u}}
\def\vb{\bar{v}}

\DeclareMathAlphabet{\mathpzc}{OT1}{pzc}{m}{it}

\begin{document}
%


\title{Vacuum Axisymmetric Gravitational Collapse revisited: preliminary investigation}

\author{Andrzej Rostworowski}
\email{andrzej.rostworowski@uj.edu.pl}
\affiliation{Institute of Theoretical Physics, Jagiellonian University, ul. S. Łojasiewicza 11, 30-348 Krak\'ow, Poland}

\begin{abstract}
Validating the results of [A.M. Abrahams and C.R. Evans, \textit{Phys. Rev. Lett.}, 70:2980–2983] poses a numerical challenge and has been inspiring a lot of research. We join these efforts and present our first steps to achieve this goal: we discuss a formulation of the Einstein equations for a vacuum axisymmetric spacetime with vanishing twist in spherical-polar coordinates, its linearised approximation, and identify some problems in achieving numerically stable evolution at the threshold of a black hole formation.  
\end{abstract}
\maketitle

\section{Introduction}
Soon after Choptuik's seminal discovery of critical phenomena in spherically symmetric gravitational collapse of a massless scalar field \cite{Choptuik:1992jv}, Abrahams and Evans (hereafter AE) reported analogous results, suggesting \textit{the universality} of a black hole (BH) mass scaling at the threshold of BH formation and existence of \textit{the universal} discretly self-similar critical solution in axisymmetric gravitational collapse of polar gravitational waves (GW) \cite{Abrahams:1993wa, Abrahams:1994nw}. Although over 30 years has passed since the publication of \cite{Abrahams:1993wa}, witnessing enormous progress in computational power and a lot of research inspired by AE findings, no one, to the best of our knowledge, was able to reproduce AE results in their original settings. Thus two questions do arise: first, are AE results reproducible and second (in case of the positive answer), is the behaviour at the BH formation threshold in axisymmetric collapse of genuine gravitational waves qualitatively the same as in the spherically symmetric model with a scalar field? The most up-to-date answer to the second question is ''probably not'' as a recent comparison of the results of three independent codes from three independent groups shows \cite{Baumgarte:2023tdh}. In particular, as the authors of \cite{Baumgarte:2023tdh} show, the behaviour at the threshold seems to be initial data dependent casting doubts on universality suggested by \cite{Abrahams:1993wa, Abrahams:1994nw} and making the criticality beyond spherical symmetry far more complicated than initially expected (for a nice summary of earlier attempts on validating AE results see \cite{Hilditch:2017dnw}). Still, the first question about reproducing AE numerical results remains valid, and worth answering. Moreover, in spite of strong evidence presented in \cite{Baumgarte:2023tdh} against universality at the threshold of BH formation, it cannot be completely ruled out, as one can always argue that the lack of observing universality is due to insufficiently precise fine tuning to the critical solution. Also, it seems that the careful use and distinction between extrinsic curvatures and extrinsic curvatures scaled by a proper power of a conformal factor (\textit{hatted} extrinsic curvatures in AE notation), present in \cite{Abrahams:1992ib,Abrahams:1993wa,Abrahams:1994nw} has been forgotten in later works \cite{Garfinkle:2000hd,Choptuik:2003as}, then rediscovered and carefully commented on by Rinne \cite{Rinne:2008tk} and then surprisingly forgotten again in \cite{Khirnov:2022rgw, Ledvinka:2021rve}. Therefore, in our opinion, going back to AE formulation may pay off. Also, the original AE setting seems to be well suited to implement higher order metric perturbations \cite{Rostworowski:2017ruj}, thus offering some theoretical advantages. 

We should also stress a long history behind \cite{Abrahams:1993wa, Abrahams:1994nw} that started with the Evans' PhD thesis \cite{Evans:1984} and then continued through linearised approximation \cite{Abrahams:1988vr}, and finally the evolution given by the Einstein equations, with dispersion for small initial data and gravitational collapse for large initial data \cite{Abrahams:1992ib}. Thus AE were well equipped to study this problem and the results presented in \cite{Abrahams:1993wa} were based on solid grounds. On the other hand it is impossible to reproduce the AE results from the paper \cite{Abrahams:1994nw}, simply because, as was pointed out in \cite{Khirnov:2022rgw}, their formula for the seed function generating initial data profiles (Eq.(6) in \cite{Abrahams:1994nw}) does not agree with what is present on their Fig. 1. Moreover this formula with its numerical constant and the value of the amplitude $a\approx1$ seems to be an order of magnitude above the threshold of BH formation, see below.

With all this in mind, we have constructed a fully constrained pseudospectral code based on AE set of equations. It performs well for small initial data that disperse, and large initial data that collapse promptly, but it is still facing some problems for the initial data close to the collapse threshold that require relatively long stable evolution in the nonlinear regime. We hope to overcome these problems in the future, with a proper filtering techniques or a better numerical treatment of the most disturbing terms in the equations, but in this paper we make a step back and focus mostly on the linearised problem that admits exact solutions. We present the evidence that in this case, the code that solves the system of linearised equations, but otherwise is based on exactly the same numerical principles as a fully nonlinear code, performs very well. The linearised solution to the AE system of equations was constructed in \cite{Abrahams:1988vr}, but we find this paper rather cumbersome and present an independent derivation based on our approach \cite{Rostworowski:2017ruj} that is quite straightforward and moreover is perfectly well suited to extend linearised approximation to higher orders - the problem that we postpone for a future work.

The paper is organised as follows. In Sec.\ref{Setting} we carefully discuss the AE ansatz for the line element and the resulting system of the Einstein equations. In Sec.\ref{Linearisation} we present in detail our derivation of the linearised approximation for the quadrupole ($\ell=2$) gravitational waves and we comment on what we believe was AE treatment of the initial data since that problem seemed to cause some confusion in the past \cite{Khirnov:2022rgw,Ledvinka:2021rve}. In Sec.\ref{Numerics} we describe the numerical techniques that we use to solve both the full system of the Einstein equations and its linearised version and present the performance of the code in solving the set of linearised equations by computing the $L^2$-norms of differences between numerical and analytical solutions. Finally, we conclude in Sec.\ref{Conclusions} and give an outlook for future work.

\section{Setting of the problem}
\label{Setting}
The original AE setting is based on quasi-isotropic spherical-polar coordinates with a simplifying assumption of the Killing vector (generating axial symmetry) being hypersurface-orthogonal (the so called \textit{twist-free} ansatz), resulting in the line element of the form
\begin{widetext}
\begin{equation}
\label{eq:LineElement}
ds^2= -\alpha^2 dt^2 + \phi^4 \left[ \left( e^{2 \eta/3} \left(dr+\br dt\right)^2 \right) + e^{2 \eta/3} r^2 \left(\frac{du}{\sqrt{1+u^2}} + \bu dt\right)^2 + e^{-4 \eta/3} r^2 \left(1-u^2\right) d\varphi^2 \right] \, ,
\end{equation}
where $0\leq r < \infty$ is the radial coordinate and $1 \geq u= \cos \theta \geq -1$ is the angular coordinate (that we use instead of the polar angle $\theta$). All five metric functions ($\eta$, $\phi$, $\alpha$, $\br$, $\bu$) are functions of $(t,r,u)$ coordinates. In particular, this ansatz encompasses the Schwarzschild BH (in isotropic coordinates):
\begin{equation}
\label{eq:SchwarzschildLineElement}
ds^2= -\left( \frac{1-M/(2r)}{1+M/(2r)} \right)^2 dt^2 + \left( 1+M/(2r) \right)^4 \left( dr^2 + r^2 \frac{du^2}{1+u^2} +r^2 \left(1-u^2\right) d\varphi^2 \right) \,,
\end{equation}
\end{widetext}
but it does not encompass the Kerr BH, which is due to the twist-free property. In other words, referring to the jargon of metric perturbations,	 the line element (\ref{eq:LineElement}) is a consistent truncation to the polar sector of perturbations, with $\eta$ being the only one (polar) gravitational degree of freedom (instead of two gravitational degrees of freedom present in the general ansatz). Although polar (party-even) and axial (parity-odd) perturbations do decouple at the linear level they get coupled at higher orders of perturbation expansion around the background solution (Minkowski in the present case). However, at axial symmetry, polar linear perturbation excites only polar perturbations at higher orders (there is no consistent truncation for axial perturbations: a linear axial perturbation would excite polar perturbations at even orders and axial perturbations at odd orders of a perturbation expansion). It follows that a Schwarzschild BH is a natural candidate for the end-state of the evolution of ``large'' (collapsing) initial data.

The integration strategy for the ansatz based on (\ref{eq:LineElement}), supplemented with the maximal slicing condition, is nicely summarized in the Appendix F of \cite{Baumgarte_Shapiro_2010} (see also \cite{Rinne:2008tk} for the analogous treatment in cylindrical-polar coordinates). Here we just adopt this discussion to our angular coordinate choice. To partially fix the gauge, we assume the maximal slicing condition: $\krr+\kuu+\kff=0$, with the standard definition of extrinsic curvatures:
\begin{align}
\label{eq:ExtrinsicCurvature}
&K^{\mu}_{\,\,\nu} = g^{\mu \lambda} K_{\lambda \nu}\,, \qquad K_{\lambda\nu} = \mathcal{L}_n g^{\perp}_{\lambda\nu} \,,
\end{align}
where
\begin{align}
n_{\mu}=(-\alpha,0,0,0)\,,\qquad g^{\perp}_{\mu\nu}=g_{\mu\nu} + n_{\mu}n_{\nu}\,,
\end{align}
$\mathcal{L}_n$ denotes the Lie derivative along $n$ and $g_{\mu\nu}$, $g^{\mu\nu}$ are the metric and its inverse, corresponding to the line element (\ref{eq:LineElement}). It is useful to introduce the following (scaled) \textit{numerical} variables:
\begin{align}
\tildeta&=\frac{\eta}{1-u^2}, \quad \psi=\phi \, e^{-\eta/3} , \quad \chi=\alpha\,\psi \,, 
\\
\tildbr &= \frac{\br}{r}, \quad \tildkrr=\phi^6\, \krr, 
\quad \tildll = \phi^6\, \frac{\lambda}{1-u^2}  \,,
\\
\tildbu &= \frac{\bu}{\sqrt{1-u^2}}, \qquad \tildkru = \phi^6\, \frac{\kru}{r} \,,
\end{align}
where $\lambda \equiv \krr + 2 \kff$ is a convenient linear combination of extrinsic curvatures. For the quadrupole, $\ell=2$, initial data that we discuss below, the functions $\tildbu$ and $\tildkru$ are odd in $u$, while all other functions are even in $u$.

From the definitions of extrinsic curvatures (\ref{eq:ExtrinsicCurvature}), maximal slicing condition and the Einstein equations $R_ {\mu \nu} = 0$,
we get a closed system of equations allowing for fully constrained numerical evolution. In particular we get the following evolution equations:
\begin{equation}
\label{eq:etaDot}
\pdt \tildeta=\br \pdr \tildeta + \tildbu \pdu \eta + \pdu \tildbu +\alpha \frac{\tildll}{\phi^6} \,,
\end{equation}
\begin{widetext}
\begin{align}
&\pdt \tildkru = \tildbr \, r\pdr \tildkru + \tildbu \left(1-u^2\right) \pdu \tildkru + \left[2 \left(1-u^2\right) \pdu \tildbu -4u \tildbu + 3 \tildbr\right] \tildkru + \frac12 \left[ 3 \tildkrr + \left(1-u^2\right) \tildll \right] \pdu \tildbr
\nonumber\\
&+\frac{1}{r}\left\{ \psi \left[ (\pdu \eta) (\pdr \chi) + (\pdr \eta) (\pdu \chi) \right] + \chi \left[(\pdu \eta) (\pdr \psi) + (\pdr \eta) (\pdu \psi)\right] + 3 \left[ (\pdu \psi)(\pdr \chi) + (\pdr \psi) (\pdu \chi) \right] \right\}
\nonumber\\
& - \psi \, \pdru \frac{\chinum}{r} - \chi \, \pdru \frac {\psinum} {r} + \psi \, \chi \left[ \frac{1}{r^2} \pdu \eta  - \frac{u}{r} \pdr \tildeta\right] \,.
\label{eq:KruDot}
\end{align}
We can choose the initial profiles for $\tildeta$ and $\tildkru$ as free data for this pair of the two \textit{unconstrained} dynamical variables. Given their initial values on some initial time-slice $t$ all remaining extrinsic curvatures and metric functions can be set on this slice solving (in sequence) the following equations:
\begin{align}
\label{eq:K1}
&\pdu \tildkrr + \left(\pdu \eta\right) \tildkrr + \left(1-u^2\right) \pdu \tildll - \left[ 4u + \left(1-u^2\right) \pdu \eta \right] \tildll = 2 \left[r \pdr \tildkru + 3 \tildkru \right] \,,
\\
\label{eq:K2}
&r \pdr \tildkrr + \left( 3 - \frac12 \, r \pdr \eta \right) \tildkrr + \frac12  \left(1-u^2\right) r \left(\pdr \eta\right) \tildll = - \pdu \left[\left(1-u^2\right) \tildkru \right]  \,,
\end{align}
then
\begin{align}
&r \pdr \left[r \pdr \psinum \right] + r \pdr\psinum + \pdu \left[\left(1-u^2\right) \pdu \psinum \right]  + \frac{1}{4}\left[ r \pdr \left(r \pdr \eta \right) + \left(1-u^2\right) \pduu \eta - u \pdu \eta\right]\psinum
\nonumber\\
=&-\frac{1}{4}\left\{ r \pdr \left(r \pdr \eta \right) + \left(1-u^2\right) \pduu \eta - u \pdu \eta + \frac{r^2 e^{-2 \eta}}{4 \left[1+\psinum\right]^7} \left[
 \left(\left(1-u^2\right) \tildll\right)^2 + 3 \left(\tildkrr\right)^2 + 4\left(1-u^2\right) \left(\tildkru\right)^2 
\right]\right\} \,,
\label{eq:psinum}
\end{align}
then
\begin{align}
&r \pdr \left[r \pdr \chinum \right] +r \pdr\chinum + \pdu \left[\left(1-u^2\right) \pdu \chinum \right] 
\nonumber\\
&+ \frac{1}{4}\left\{ r \pdr \left(r \pdr \eta \right) + \left(1-u^2\right) \pduu \eta - u \pdu \eta - \frac{7}{\psi^8}\frac{r^2}{4}e^{-2 \eta} \left[
 \left(\left(1-u^2\right) \tildll\right)^2 + 3 \left(\tildkrr\right)^2 + 4\left(1-u^2\right) \left(\tildkru\right)^2 
\right]\right\}\chinum
\nonumber\\
=&-\frac{1}{4}\left\{ r \pdr \left(r \pdr \eta \right) + \left(1-u^2\right) \pduu \eta - u \pdu \eta - \frac{7}{\psi^8}\frac{r^2}{4}e^{-2 \eta} \left[ \left(\left(1-u^2\right) \tildll\right)^2 + 3 \left(\tildkrr\right)^2 + 4\left(1-u^2\right) \left(\tildkru\right)^2 
\right]\right\} \,,
\label{eq:chinum}
\end{align}
\end{widetext}
and then finally
\begin{align}
\label{eq:shifts1}
r \pdr \tildbr - \left(1-u^2\right) \pdu \tildbu + u \tildbu &=\frac{\alpha}{2} \frac {\left(1-u^2\right) \tildll + 3 \tildkrr}{\phi^6} \,,
\\
\label{eq:shifts2}
\pdu \tildbr + r \pdr \tildbu &= 2 \alpha \frac {\tildkru}{\phi^6} \,.
\end{align}
Once all equations (\ref{eq:K1}-\ref{eq:shifts2}) are solved we can use the evolution equations (\ref{eq:etaDot},\ref{eq:KruDot}) to step in time to the next time-slice $t+dt$. We note that Eq.(\ref{eq:psinum}) is the only nonlinear equation, while Eqs.(\ref{eq:K1},\ref{eq:K2},\ref{eq:chinum}-\ref{eq:shifts2}) are linear (for the unknowns being solved for) and straightforward to solve. Since for the flat Minkowski spacetime $\psi \equiv 1 \equiv \chi$, it is much better numerically to solve for $(\psi-1)$ and $(\chi-1)$ instead of solving for $\psi$ and $\chi$ themselves. The partial decoupling of equations in the system (\ref{eq:K1}-\ref{eq:shifts2}) comes from the maximal slicing condition and evolving in time for $\phi^6\,\kru$ and \textit{not} $\kru$ itself. We stress again that it is crucial to introduce the scaled extrinsic curvatures (as AE did), as only then Eq.(\ref{eq:psinum}) decouples form Eqs.(\ref{eq:K1},\ref{eq:K2}) and moreover we can expect the Hamiltonian constraint, Eq.(\ref{eq:psinum}), to have a unique solution, see \cite{Rinne:2008tk} for a thorough discussion. 

A remark about AE initial data is in order here, since it seems that this issue caused some confusion in the past. At the linear level $\phi^6\, K^i_{\,\,j} = K^i_{\,\,j}$. Thus, in view of a clear advantage of evolvng in time $\phi^6 \kru$, instead of $\kru$ itself, it is quite natural to expect that AE took Eq.(4) in \cite{Abrahams:1992ib} (and Eq.(5) in \cite{Abrahams:1994nw} respectively) as the initial profile for $\phi^6\,\kru/r = \tildkru$ in our notation, and \textit{not} for $\kru/r$ itself. In other words the linear approximation for $\tildkru$ is $\kru/r$ and this is what provides the initial data for their evolution of the full set of the Einstein equations. This should have been stated more clearly in \cite{Abrahams:1992ib,Abrahams:1993wa} but it can be read off from the context, when both these papers are studied together. It seems that this fact was overlooked in \cite{Khirnov:2022rgw, Ledvinka:2021rve} and led to the problem of non-unique solutions of the Hamiltonian constraint while solving the coupled system (\ref{eq:K1}-\ref{eq:psinum}) for not scaled component of extrinsic curvature and $\psi$ (see again the discussion in \cite{Rinne:2008tk}). There may be also some further confusion about AE initial data from \cite{Abrahams:1992ib,Abrahams:1993wa} caused by the fact that AE do not specify the numerical values of the dimensional parameters $\lambda$ and $r_0$ (characterizing the width and the position of their initial data wave packet, see the discussion above Eq.(2) in \cite{Abrahams:1992ib}) that they use in their numerical simulation. Although we do not find it to be a good practice in a numerical work, this in principle should not cause any confusion since $\lambda$ can be considered to be the only dimensional parameter in the the problem, and in their plots, AE scale the time coordinate with the corresponding values of ADM mass set by the initial data (the dimensionless value $r_0/\lambda=20$(?) can then be guessed from the inset plot in the Fig.1 of \cite{Abrahams:1994nw}).

To close this section, we note that although AE used a partially-free evolution scheme, i.e. they traded the linear system of first order PDEs (momentum constraints) (\ref{eq:K1},\ref{eq:K2}) for the evolution equations for the corresponding components of the extrinsic curvatures (see Eqs.(3,4) in \cite{Abrahams:1993wa}), we opt for a fully constrained scheme, as given by the set of Eqs. (\ref{eq:K1}-\ref{eq:shifts2}), since it is expected to have much better stability properties. On the other hand, in such a fully constrained scheme, one is in principle free to single out the extrinsic curvature component to be evolved in time (in a pair with $\eta$) and this freedom should be used for the numerical advantage. Here, after experimenting with this choice, we pick up $\tildkru$ to be evolved dynamically, as only with this choice we have managed to achieve spectral accuracy when solving momentum constraints (\ref{eq:K1},\ref{eq:K2}).

\section{Linear approximation}
\label{Linearisation}
Checking the evolution provided by a numerical code against analytically known, static or linearized solutions of the Einstein equations is the first step in numerical code validation. In particular the numerical method should provide stable time evolution for the linearized Einstein equations themselves. Thus, to test the numerical algorithm it is very useful to have exact solutions of the linearized Einstein equations around Minkowski spacetime. The general solution to the linearized Einstein equations on Minkowski background can be easily found using \cite{Rostworowski:2017ruj}. In particular, expanding
\begin{equation}
\label{eq:pertExpansion}
g_{\mu\nu} = \eta_{\mu\nu} + \epsilon \, h_{\mu\nu} + \mathcal{O}\left( \epsilon^2 \right) \,,
\end{equation}
where $ \eta_{\mu\nu}$ is the Minkowski metric, we wish to satisfy the Einstein equations up to linear order in $\epsilon$. $h_{\mu\nu}$ is determined up to (linear) gauge transformations, i.e. $h_{\mu\nu}$ and $h_{\mu\nu} - \bar{\nabla}_\mu \xi_\nu - \bar{\nabla}_\nu \xi_\mu$ with arbitrary gauge vector $\xi_\mu$ represent the same solution (here $\bar{\nabla}$ is the background Minkowski space metric connection). In axial symmetry, it is useful to expand polar components of the metric perturbations (in general: any symmetric tensor) $h_{\mu\nu}$ as follows \cite{Nollert:1999ji,Rostworowski:2017ruj}:
\begin{align}
\label{eq:polarT_ab}
h_{ab}(t,r,u) = & \sum_{0 \leq \ell} \hl_{ab} (t,r) P_{\ell}(u), \quad a,b=t\,,r \, , 
\\
\label{eq:polarT_a2}
h_{au}(t,r,u) = & \sum_{1 \leq \ell} \hl_{au} (t,r) \pdu P_{\ell}(u), \quad a=t\,,r \, , 
\end{align}
\begin{align}
&\frac{1}{2}\left( \left( 1-u^2 \right) h_{uu}(t,r,u) + \frac{h_{\varphi\varphi}(t,r,u)}{1-u^2}\right) 
\nonumber\\
&=  \sum_{0 \leq \ell} \hl_{+} (t,r) P_{\ell}(u) \, ,
\label{eq:polarT+}
\\
&\frac{1}{2}\left( \left( 1-u^2 \right) h_{uu}(t,r,u) - \frac{h_{\varphi\varphi}(t,r,u)}{1-u^2}\right)  
\nonumber\\
&= \sum_{2 \leq \ell} \hl_{-} (t,r) \left(-\ell(\ell+1) P_{\ell}(u) + 2 u \pdu P_{\ell}(u)\right) \, ,
\label{eq:polarT-}
\end{align}
and to expand polar components of a (gauge) vector as
\begin{align}
\label{eq:polarV_a}
\xi_{a}(t,r,u) = & \sum_{0 \leq \ell}\xil_a (t,r) P_{\ell}(u), \quad a = t\,,r \, , 
\end{align}
\begin{align}
\label{eq:polarV_2}
\xi_{u}(t,r,u) = & \sum_{1 \leq \ell} \xil_u (t,r) \pdu P_{\ell}(u) \, ,
\end{align}
where $P_{\ell}(u)$ are Legendre polynomials. Then, the linearized Einstein equations are solved with \footnote{see Eqs. (22-28), (38-41) and (36,37) in \cite{Rostworowski:2017ruj}}:
\begin{equation}
\end{equation}
\begin{align}
\label{eq:htt}
\hl_{tt} &= \pdr \left( r \pdr \Phil  \right)  - \frac{\ell(\ell+1)}{2 r} \Phil + 2 \pdt \xil_t \,,
\\
\hl_{tr} &=  \pdr \left( r \pdt \Phil  \right) + \pdr \xil_t + \pdt \xil_r \,,
\\
\hl_{tu} &= \pdt \xil_u + \xil_t  \,,
\\
\hl_{rr} &= \pdr \left( r \pdr \Phil  \right)  - \frac{\ell(\ell+1)}{2 r} \Phil + 2 \pdr \xil_r \,, 
\\
\hl_{ru} &= \xil_r - \frac2r \xil_u + \pdr \xil_u \,,
\\
\hl_{+} &= r^2 \left(\pdr \Phil + \frac{\ell(\ell+1)}{2 r} \Phil \right) + 2 r \xil_r - \ell(\ell+1) \xil_u \,,
\\
\label{eq:h-}
\hl_{-} &= \xil_u \,,
\end{align}
where $\Phil=\Phil(t,r)$ are (polar) master scalars that for $\ell \geq 2$ satisfy a scalar wave equation on Minkowski background: 
\begin{equation}
\label{eq:masterEq}
\pdtt \Phil - \pdrr \Phil + \frac{\ell(\ell+1)}{r^2} \Phil = 0\,.
\end{equation}
The general solution of the master equation (\ref{eq:masterEq}) reads
\begin{equation}
\label{eq:masterScalar}
\Phil= \sum_{0\leq k \leq \ell} \left[ \frac{(2\ell - k)!}{k! \, (\ell-k)!} \frac{f_\ell^{(k)}(\ub)}{(2r)^{l-k}} + (-1)^{\ell+k} \frac{(2\ell - k)!}{k! \, (\ell-k)!} \frac{g_\ell^{(k)}(\vb)}{(2r)^{l-k}} \right] \,,
\end{equation}
where $\ub=t-r$, $\vb=t+r$ are outgoing and ingoing null coordinates, $f_\ell$ and $g_\ell$ are arbitrary profiles, $f_\ell^{(k)}(x) \equiv d^k/dx^k\, f_\ell(x)$, and similarly for $g_\ell^{(k)}(x)$. The regularity condition, for master scalars to be regular in the center ($r=0$), yields the $g_\ell=-f_\ell$ conditions. For Minkowski (linear) perturbations, $\ell=0,1$ multipoles in (\ref{eq:pertExpansion}) correspond to a pure gauge thus for $\ell=0,\,1$, $\Phil$ are identically zero. 
Gauge vector components in (\ref{eq:htt}-\ref{eq:h-}) can be adjusted in order to comply with the form of the line element (\ref{eq:LineElement}), regularity conditions for metric functions, maximal slicing condition, and gauge choice  for the lapse function. In particular, since the $drdu$ term is absent in the line element (\ref{eq:LineElement}) we must have
\begin{equation}
\label{eq:xilrGaugeCondition}
\xil_r = -\pdr \xil_u + \frac2r \xil_u \,, \qquad \ell \geq 1 \,.
\end{equation}
\subsection{$\ell=2$ quadrupolar waves}
\noindent
Here we focus on the case of quadrupolar waves, $\ell=2$. In particular, expanding the metric functions
\begin{align}
\eta &= \epsilon\, q_1  + \mathcal{O}\left( \epsilon^2 \right) \,, \quad 
\\
\phi &=  1+ \epsilon\, q_2  + \mathcal{O}\left( \epsilon^2 \right) \,, \quad
\\
\alpha &=  1+ \epsilon\, q_3  + \mathcal{O}\left( \epsilon^2 \right) \,, \quad
\\
\br &= \epsilon\, q_4  + \mathcal{O}\left( \epsilon^2 \right) \,, \quad
\\
\bu &= \epsilon\, q_5  + \mathcal{O}\left( \epsilon^2 \right) \, 
\end{align}
and equating the metric obtained for the line element (\ref{eq:LineElement}) expanded up to the first order in $\epsilon$ with (\ref{eq:pertExpansion}-\ref{eq:polarT-}), with the sums in (\ref{eq:polarT_ab}-\ref{eq:polarT-}) truncated at $\ell=2$ (it is necessary to include $\ell=0$ gauge degrees of freedom in the sums), we get \footnote{first, from $tt$, $tr$, $tu$, $uu$ and $\varphi\varphi$ components of the metric we get Eqs. (\ref{eq:q1}-\ref{eq:q5}) and then, from the $rr$ component, we get Eqs. (\ref{eq:hl2rr},\ref{eq:hl0rr}).} 
\begin{align} 
\label{eq:q1}
q_1 &= \frac{3}{r^2} \left( 1- u^2 \right) \hltwo_{-} \,,
\end{align}
\begin{align} 
q_2 &= \frac{1}{8 r^2} \left( - \left( 1 - 3u^2 \right) \hltwo_{+} + 2 \left( 1- u^2 \right) \hltwo_{-} + 2 \hlzero_{+} \right) \,,
\\
q_3 &= \frac{1}{4} \left( \left( 1 - 3u^2 \right) \hltwo_{tt} - 2 \hlzero_{tt} \right) \,,
\\
q_4 &= -\frac{1}{2} \left( 1 - 3u^2 \right) \hltwo_{tr} + \hlzero_{tr}\,,
\\
\label{eq:q5}
q_5 &= \frac{3}{r^2} u \sqrt{1- u^2} \hltwo_{tu} 
\end{align}
and
\begin{align} 
\label{eq:hl2rr}
\hltwo_{r,r} &= \frac1{r^2} \left( \hltwo_{+} - 2 \hltwo_{-} \right) \,,
\\
\label{eq:hl0rr}
\hlzero_{r,r} &= \frac1{r^2} \left( \hlzero_{+} + 2 \hltwo_{-} \right) \,,
\end{align}
which translates into (cf. Eq. (\ref{eq:xilrGaugeCondition}))
\begin{align}
\label{eq:xiltwou}
2r \pdrr \xiltwo_u - 6 \pdr \xiltwo_u &= r^2 \pdrr \Philtwo - 6 \Philtwo \,,
\\
r ^2 \pdr \xilzero_r  - r \xilzero_r&= \xiltwo_u \,.
\end{align}
Integration of these equations yields \footnote{After putting the Eq.(\ref{eq:xiltwou}) into a form 
\[
\pdr \left( \frac1{r^3} \pdr \xiltwo_u \right) = \pdr \left( \frac{\pdr \Philtwo}{2r^2} + \frac{\Philtwo}{r^3} \right) \,,
\]
which is suitable for taking its 1st radial integral, we use the explicit form of $\Philtwo$:
\[
\Philtwo = f''_2(\ub) + g''_2(\vb) + 3 \frac{ f'_2(\ub) - g'_2(\vb)}{r}+ 3 \frac{ f_2(\ub) + g_2(\vb)}{r^2} \, ,
\]
cf. Eq. (\ref{eq:masterScalar}).}
\begin{widetext}
\begin{align}
\xiltwo_u  &= \frac12 \left( r f_2^{(2)}(\ub) + r g_2^{(2)}(\vb) + 2 f_2^{(1)}(\ub) - 2 g_2^{(1)}(\vb) + \cltwo_{u,1}(t) + r^4 \cltwo_{u,2}(t) \right)
\\
\xilzero_r  &= -\frac1{2r} \left( f_2^{(1)}(\ub) - g_2^{(1)}(\vb) + \cltwo_{u,1}(t) \right) + r \clzero_r(t) \,.
\end{align}
We put $\cltwo_{u,2}(t) \equiv 0 \equiv \clzero_r(t)$ to assure the correct asymptotic behavior (otherwise $h_{\mu\nu}$ would not be explicitly asymptotically flat). Thus we get
\begin{align}
\label{eq:etaLin}
\eta &= \epsilon\, \left( 1 - u^2 \right) \frac32 \left( \frac{ f_2^{(2)}(\ub) + g_2^{(2)}(\vb)}{r} + 2 \frac{f_2^{(1)}(\ub) - g_2^{(1)}(\vb)}{r^2} + 2 \frac{\cltwo_{u,1}(t)}{r^2} \right)  + \mathcal{O}\left( \epsilon^2 \right) \,, 
\\
\label{eq:phiLin}
\phi-1 &=  \epsilon\, \frac18 \left[ \left( 1 - u^2 \right) \left( \frac{ f_2^{(2)}(\ub) + g_2^{(2)}(\vb)}{r} - \frac{f_2^{(1)}(\ub) - g_2^{(1)}(\vb)}{r^2} \right) - \left( 3 - 9u^2 \right) \frac{f_2(\ub) + g_2(\vb)}{r^3} + \left( 2 - 8u^2 \right) \frac{\cltwo_{u,1}(t)}{r^2} \right]  
\nonumber\\
&+ \mathcal{O}\left( \epsilon^2 \right) \,.
\end{align}
\end{widetext}
The regularity condition in the center for the metric function $\eta$ (taking into account $g_2=-f_2$ condition) requires:
\begin{equation}
\label{eq:gaugeRegularity}
\cltwo_{u,1}(t) = - 2 f_\ell^{(1)}(t) \,.
\end{equation}  
We note that (up to a convention (3/2) factor) our Eq. (\ref{eq:etaLin}), after setting $f_2 = 0 = \cltwo_{u,1}$,  $g_2=I$, coincides with the Eq. (3) in \cite{Abrahams:1992ib} (also taken as the initial profile for $\eta$ in \cite{Abrahams:1993wa}), and after setting $g_2=-f_2=-I$, it coincides with Eq.(4) in \cite{Abrahams:1994nw}, with a factor 2 missing in front of $b_2(t)$ in this formula (see the text below Eq.(4) in \cite{Abrahams:1994nw} for the definition of $b_2(t)$, that corresponds to our condition (\ref{eq:gaugeRegularity})). The time components of the gauge vectors (entering $\alpha$, $\br$ and $\bu$ metric functions) are set by the maximal slicing condition. A rather involved integration of this gauge condition yields:
\begin{widetext}
\begin{align}
\xiltwo_t  &= -\left( r \frac{ f_\ell^{(3)}(\ub) + g_\ell^{(3)}(\vb)} 2 + f_\ell^{(2)}(\ub) - g_\ell^{(2)}(\vb) + 3 \frac{f_\ell^{(1)}(\ub) + g_\ell^{(1)}(\vb)}{2r} + 3 \frac{ f_\ell(\ub) - g_\ell(\vb)}{2r^2} + 3 \frac{ f_\ell^{(-1)}(\ub) + g_\ell^{(-1)}(\vb)}{2r^3} \right) \,,
\\
\xilzero_t  &= 0 \,,
\end{align}
where we have put all integration constants to zero for our convenience, and
\begin{equation}
\label{eq:FG}
F_2(x) \equiv f_2^{(-1)}(x)= \int^x f_2(y)\, dy \,, \quad G_2(x) \equiv g_2^{(-1)}(x)= \int^x g_2(y)\, dy \,,
\end{equation}
with $F_2=-G_2$ condition, which again follows from the requirement of regularity at $r=0$ (thus arbitrary integration constants allowed by the definitions (\ref{eq:FG}) drop out). Collecting these results together we finally get:
\begin{align}
\label{eq:kruLin}
\frac{\kru}{r} &=  \epsilon\, 3 u \left( \frac{ f_2^{(2)}(\ub) - g_2^{(2)}(\vb)}{r^2} + 3 \frac{f_2^{(1)}(\ub) + g_2^{(1)}(\vb)}{r^3} + 6 \frac{f_2(\ub) - g_2(\vb)}{r^4} + 6 \frac{f_2^{(-1)}(\ub) + g_2^{(-1)}(\vb)}{r^5}\right) + \mathcal{O}\left( \epsilon^2 \right) \,.
\end{align}
\end{widetext} 
We note that taking $F_2=-G_2$ to be even we get time-symmetric initial data (distorted spatial metric and extrinsic curvatures and shifts being zero at the moment of time symmetry $t=0$), and  taking $F_2=-G_2$ to be odd we get time-antisymmetric initial data (Euclidean spatial metric and nonzero values for extrinsic curvatures and shifts at the moment of time asymmetry $t=0$). We also note that (again, up to a convention (3/2) factor) our Eq. (\ref{eq:kruLin}) after setting $f_2 = 0$,  $g_2=I$ coincides with the Eq. (4) in \cite{Abrahams:1992ib} (also taken as the initial profile for $\hat{K}^r_{\,\,\theta}$ in \cite{Abrahams:1993wa}; mind the $(-\sin\theta)=du/d\theta$ factor from the angular coordinate transformation), and after setting $g_2=-f_2=-I$ it coincides with Eq.(5) in \cite{Abrahams:1994nw}, this time with an overall (-1) factor missing in front of this formula (as we have already fixed the sign convention $g_2=-f_2=-I$ when comparing our Eq. (\ref{eq:etaLin}) to the Eq.(4) in \cite{Abrahams:1994nw}).
\subsection{Definition of the linearized approximation to the spacetime mass}
Our linear approximation (\ref{eq:pertExpansion}) solves the Einstein equations up to linear order in $\epsilon$. Thus, calculating the Einstein tensor following from (\ref{eq:pertExpansion}) we will in general get
\begin{equation}
\label{eq:EMtensor}
G_{\alpha\beta}\left[ \eta_{\mu\nu} + \epsilon \, h_{\mu\nu} \right] = \epsilon^2 \, \left(- 8 \pi G \, T_{\alpha\beta} \right) + \mathcal{O} \left( \epsilon^3 \right)\,,
\end{equation}
where the coefficent of $\epsilon^2$ on the RHS of this equation can be interpreted as $(-8\pi G)$ times the energy-momentum tensor sourcing the 2nd-order metric perturbations (the minus sign comes from the fact that writing the 2nd-order perturbation equations in the form of the Einstein equations ''\textit{geometry=matter content}'', this term would be moved to the \textit{matter} side (with a sign change) to source the \textit{geometry} side of the 2nd-order metric perturbations). We can thus \textit{define} local energy density of (linear) gravitational waves as $\epsilon^2 \, T_{tt}$ from Eq. (\ref{eq:EMtensor}) and the total mass of the wave packet $\mp$ as the integral over the initial hypersurface:
\begin{align}
\mp&= \epsilon^2 \, 2 \pi \int_0^\infty r^2 dr \int_{-1}^{+1} du \, T_{tt}
\nonumber\\
&= - \frac{1}{4 G} \int_0^\infty r^2 dr \int_{-1}^{+1} du \, G_{tt} \left[ \eta_{\mu\nu} + \epsilon \, h_{\mu\nu} \right] \,.
\end{align}
For small initial data we can expect $\mp$ to be a good approximation for Arnowitt-Deser-Misner (ADM) mass $\madm$, with $\madm < \mp$ by a fraction of $\mathcal{O}\left( \mp/\sigma \right)$, where $\sigma$ is a scale of the (spatial) width of the packet. We can easily calculate $\mp$ for some model initial data:
\subsubsection{}
For the initial data studied in \cite{Abrahams:1992ib,Abrahams:1993wa}, $f_2^{(-2)}(x)= \sigma^5 \left(1 - (x/\sigma)^2 \right)^6$ for $|x|<\sigma$ and zero otherwise, we get
\begin{equation}
G \mp = \left(\epsilon \, \frac32 \, \frac{12288}{\sqrt{143/\pi}} \, \frac1{\sqrt{2\pi}} \right)^2 \sigma \,.
\end{equation}
The (3/2) factor accounts for a different convention between our Eqs. (\ref{eq:etaLin},\ref{eq:kruLin}) and the Eqs. (3,4) in \cite{Abrahams:1992ib}. The $\left(12288/\sqrt{143/\pi}\right)$ factor explains the funny choice of the constant $\kappa$ made in \cite{Abrahams:1992ib,Abrahams:1993wa}. Namely, defining
\[
\epsilon = (2/3) \kappa \, A = (2/3) \left(\sqrt{143/\pi} / 12288\right) \, A \,,
\]
we get $G \mp = \sigma A^2 / (2 \pi)$ which agrees with the Eq. (5) in \cite{Abrahams:1992ib}. These data are not smooth (the 6th derivative of $f_2^{(-2)}$ is discontinuous) thus they are not well suited for pseudo-spectral methods that we use. Thus we will not test these initial data with our code.
\subsubsection{}
For the time-antisymmetric initial data studied in \cite{Abrahams:1994nw}, $f_2^{(-2)}(x)= \sigma^5 e^{- (x/\sigma)^2}$, we get
\begin{equation}
G \mp = \left(\epsilon \, \left( \frac\pi8 \right)^{(1/4)} 9 \sqrt{7} \right)^2 \sigma \,,
\end{equation}
which shows that the funny factor $c_0 = \sqrt{225/8\pi}$ from \cite{Abrahams:1994nw} does not make much sense for this seed profile (even taking into account the (3/2) convention factor). Instead, taking $\epsilon = \left( \frac8\pi \right)^{(1/4)} \left( 1/ (9 \sqrt{7}) \right) \, A$, we get $G \mp = \sigma A^2$.
We were quite happy when we discovered the paper \cite{Abrahams:1994nw}, as the initial data resulting from this seed profile, being smooth, would be perfectly well suited to evolve with a pseudospectral code.
Unfortunately, as it was pointed out in \cite{Khirnov:2022rgw}, the initial data for this seed profile do not correspond to the initial data from the Fig.1 in \cite{Abrahams:1994nw} and no one knows which initial data were in fact evolved in this work. Thus, unfortunately, we cannot take this data to compare our code's results with those of AE. 
\subsubsection{}
\label{forTests}
To test our code we chose the time-symmetric initial data, given by a seed function $F_2(x)=f_2^{(-1)}(x)= \sigma^4 e^{- (x/\sigma)^2}$, we get
\begin{equation}
G \mp = \left(\epsilon \, \left( \frac\pi8 \right)^{(1/4)} \sqrt{63} \right)^2 \sigma \,.
\end{equation}
Taking $\epsilon = \left( \frac8\pi \right)^{(1/4)} \left( 1/ \sqrt{63} \right) \, A$, we get $G \mp = \sigma A^2$.

Following {\cite{Abrahams:1992ib}} we take the value of $G \mp / \sigma =: A^2$ as the dimensionless parameter characterizing the strength of the initial data obtained for a given seed profile $f_2=-g_2$. When evolving such data with the Einstein equations, we should see a linearized regime for $A^2 \ll 1$ and we could expect transition to collapse for $A$ of the order of one.
\section{Numerical Results}
\label{Numerics}
In the numerical algorithm that we use to solve the system (\ref{eq:etaDot}-\ref{eq:shifts2}), resulting from the Einstein equations, we compactify the radial grid with $r=L\,\tan (\pi x/2)$, with $0\leq x < 1$. All functions, that we solve for, do satisfy Dirichlet boundary conditions at $x=1$. Our pseudospectral code (to evaluate spatial derivatives in both: radial and angular coordinates) is based on \cite{doi:10.1137/1.9780898719598}. In particular we represent derivatives with Chebyshev differentiation matrices (see chapters 6,7 in \cite{doi:10.1137/1.9780898719598}), taking advantages of symmteries of the radial grid (a general property) and the angular grid (due to symmetry of initial data), see chapter 11 in \cite{doi:10.1137/1.9780898719598}. Due to using this double-copy technique on the numerical grid, neither $r=0$ nor the equator ($u=0$) belong to the numerical grid. On the other hand the axis, $u=1$, belongs to the numerical grid. After the system (\ref{eq:K1}-\ref{eq:shifts2}) is solved with a given dynamical pair of variables as an input, the time evolution is performed with the classic Runge-Kutta 4-th order algorithm (RK4). By performing a few numerical experiments we have found that we get the most stable time evolution by evolving a pair $\eta$ and $\tildkru$. Then the value of $\tildeta= \eta/\left( 1-u^2\right)$ at the axis, $u=1$, is set from de l'H\^opital rule (this value is needed in the evolution Eq. (\ref{eq:KruDot})). At the present stage we do not include any filtering of high frequencies on numerical grids.

\begin{figure}
\includegraphics[width=1.0\linewidth]{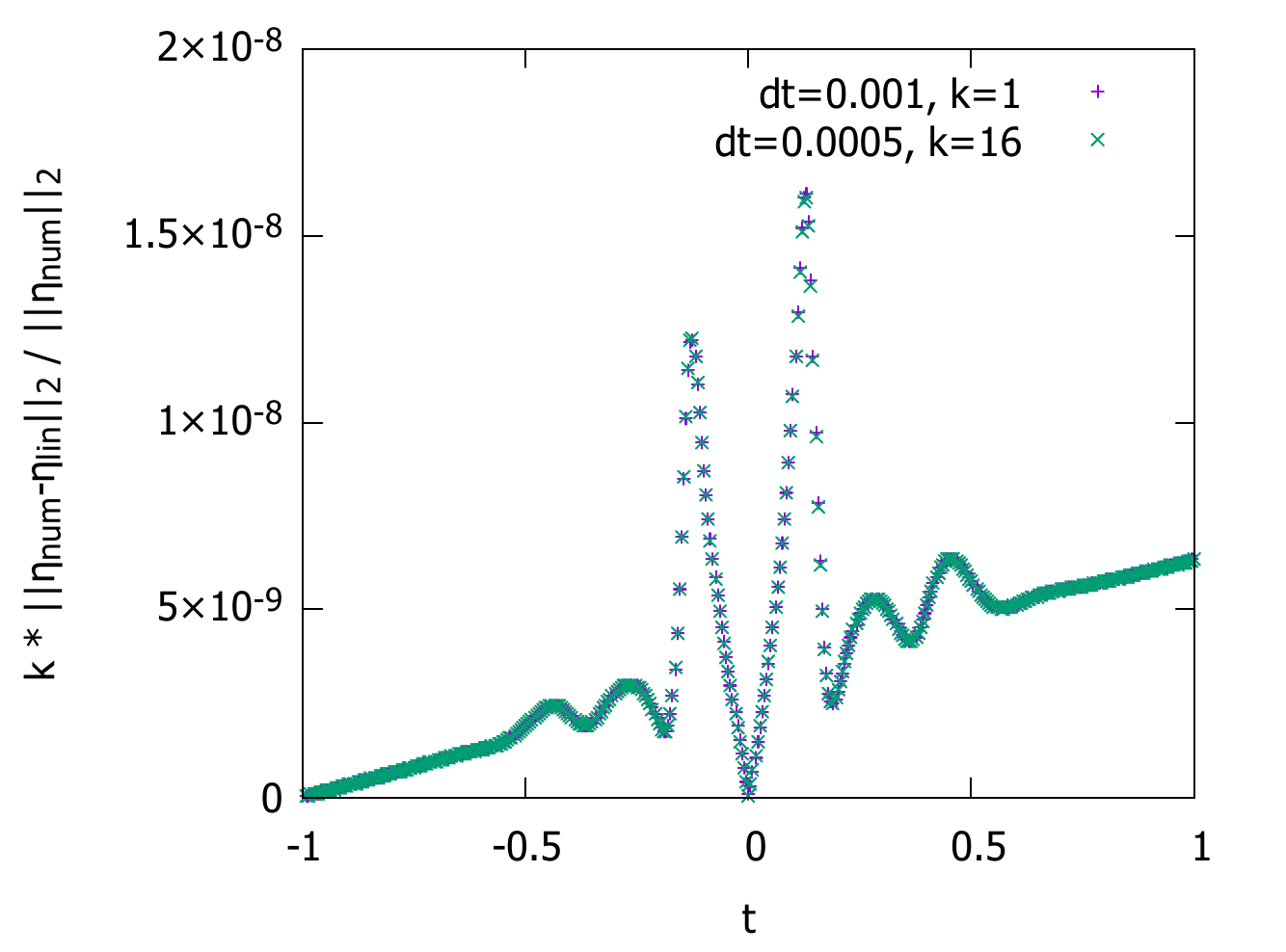}
\caption{Time evolution of $L^2$-norms of radial profiles: $\left[||\eta_{num} - \eta_{lin} ||_2 / ||\eta_{num}||_2\right](t,u)|_{u=1}$ for th RK4 scheme for time evolution with steps $dt=10^{-3}$ and $dt=0.5\times10^{-3}$. The results for the smaller time step are multiplied by a factor $k=16$ to illustrate 4th order convergence of the time evolution. The numerical grid ($n_x=401+1$, i.e. $200$ grid points in the range $0 < x < 1$) is big enough for a spectral precision to saturate (see Fig.\ref{fig:2}).}
\label{fig:1}
\end{figure}

To test our code we use time symmetric initial data generated with the seed function Sec.\ref{forTests}, with $\sigma=0.25$ and $A=1/16$ and the compactification scale $L=2$. The results of the linearised evolution are presented in the Fig.\ref{fig:1}, where we plot the $L^2$-norms of differences between numerical and analytical linearised solutions. We used the grids with $n_x=401+1$ and $n_u=15+1$ grid points on $-1 \leq x \leq 1$ and $-1 \leq u \leq 1$ grids, respectively (which translates into $200$ and $8$ grid points on the physical grids $0 < x < 1$ and $0 < u \leq 1$, respectively). The norms are taken for a radial profile along fixed polar angle $\theta=\pi/15$. The increase of the $L^2$-norms in time is entirely due to the time integration, as the ratio of the final increment between time evolutions with time-steps $dt=10^{-3}$ and $dt=0.5 \times10^{-3}$ is $16$, as expected for the 4-th order RK4 integration in time, see Fig.\ref{fig:1}.

\begin{figure}
\includegraphics[width=1.0\linewidth]{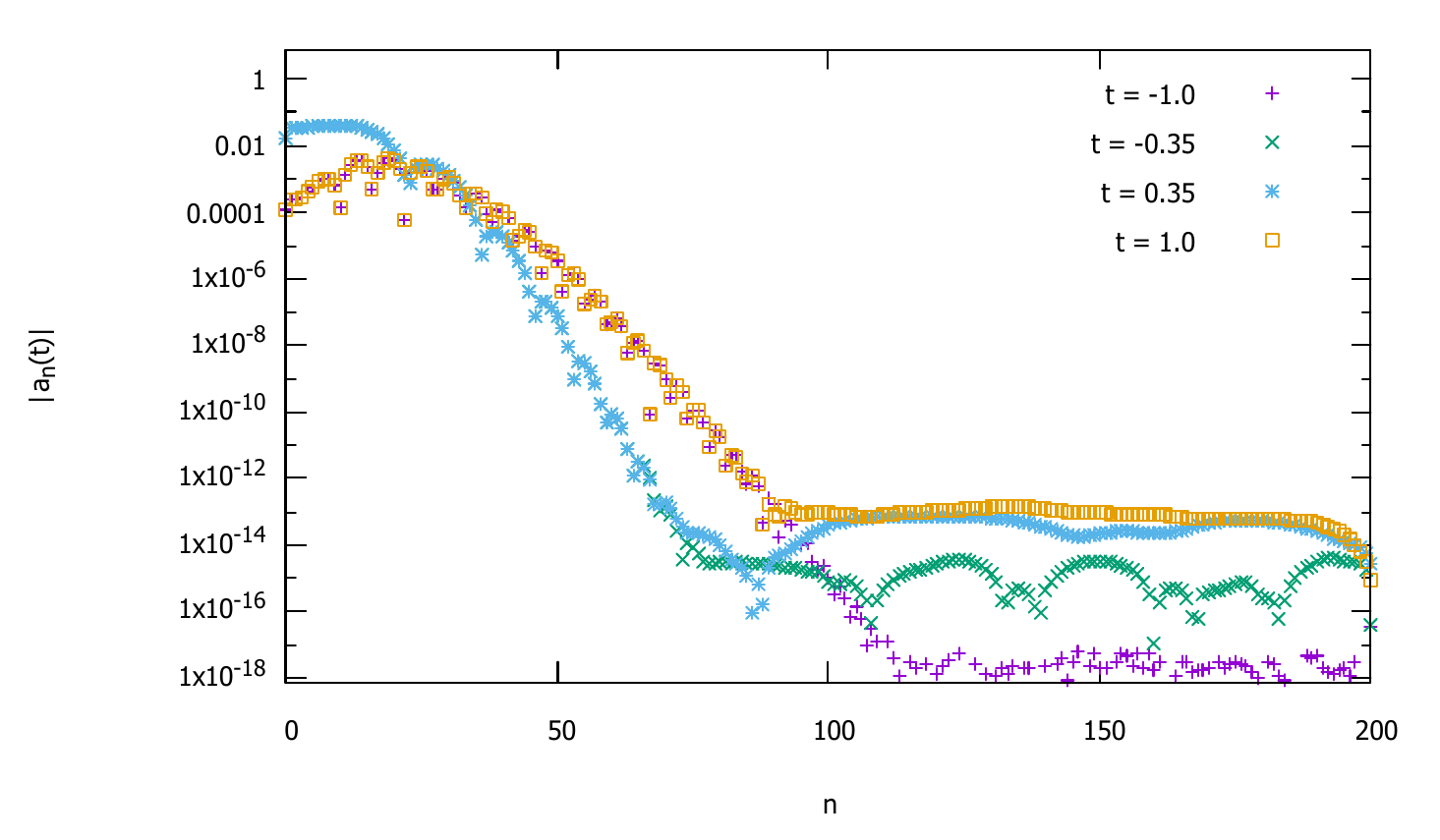}
\caption{Spectral representation of $\tildkru(t,r(x),u=1)$: absolute values of the coefficients in the expansion of $\tildkru$ into a series of Chebyshev polynomials, cf. Eq.(\ref{eq:spectral}). While the relevant parts of the spectrum for $t$ and $-t$ coincide, the plateau of (irrelevant) numerical garbage increases with time.}
\label{fig:2}
\end{figure}

In Fig.\ref{fig:2} we present what we believe is a source of numerical problems in a fully nonlinear evolution of the system (\ref{eq:etaDot}-\ref{eq:shifts2}). In particular we plot the spectral representation of the $\tildkru$ profile at the axis ($u=1$), i.e. the absolute values of the coefficients $a_n(t)$ in the series representation
\begin{equation}
\label{eq:spectral}
\tildkru(t,r(x),u=1) = \sum_n a_n(t) T_{2n}(x)
\end{equation}
for $t=-1.0\,, -0.35\,, 0.35\,, 1.0$. As can be seen on these plots the plateau of numerical noise/garbage increases with time. This efect, being harmless for linearised evolution, is rather disasterous in long-time evolution of initial data at the threshod of the collapse (i.e. for the data that do not collapse promptly), as this noise gets coupled through nonlinearities to the low-frequency part of the spectrum invalidating the time evolution (at timscale related to the ``strength'' of initial data). We identify that the terms responsible for this behaviour are the terms $\pdru \frac {\psinum} {r}$ and $\pdru \frac {\chinum} {r}$ (these two terms are actully equal in linearised equations) in the evolution Eq.(\ref{eq:KruDot}). We note that these terms are finite in the $r\rightarrow0$ limit due to the fact that for $\psinum$, $\chinum$ being scalars we have $\pdu \psinum =\mathcal{O}\left(r^2\right)$. Even though the solutions of Eqs.(\ref{eq:psinum},\ref{eq:chinum}) are spectrally accurate (for a given, spectrally acurate dynamical pair $\eta$ and $\tildkru$), the effect of dividing by $r$ and taking radial derivative is not, and it introduces some numerical garbage.

We note that figures 1,2 do not show strong $u$-dependence.

\section{Conclusions}
\label{Conclusions}
The problem of proving/disproving the validity of AE results on critical collapse of axially symmetric gravitational waves \cite{Abrahams:1993wa} still remains a numerical challenge. With this paper we join a big group of our predecessors working in this area. In particular we thoroughly discuss the AE ansatz, the role of the proper scaling of extrinsic curvature components with a proper power of a conformal factor together with, what we believe, was AE treatment of initial data. We also carefully derive the solution of the linearised Einstein equations for the line element (\ref{eq:LineElement}) supplemented with the maximal slicing condition. Finally we demonstrate a fully constrained code, combining pseudospectral approach to solve for a time-slice with RK4 evolution in time of the unconstrained dynamical pair $(\eta,\,\tildkru)$, that solves the linearised system and identify the problems we encounter while evolving the genuine set of the Einstein equations. A final success would path the way to study a bunch of interesting problems including initial data composed with higher multipoles ($\ell>2$) and/or breaking equatorial symmetry, relaxing twist free assumption, comparison with higher orders of perturbation expansion and more. We hope to witness a more complete picture of gravitational wave collapse emerging in the near future.

\section*{Acknowledgments}
The author wishes to thank Piotr Bizoń, David Hilditch, Jerzy Knopik, András László and Maciej Maliborski for useful coments and discussions. The author is also indebted to an unknown Referee for pointing out a sign error in Eqs. (57,58) in the earlier version of the manuscript and for numerous comments and remarks that significantly improved the presentation of the results. This work was supported by the National Science Centre Grant No. 2017/26/A/ST2/00530.

\bibliographystyle{plain}
\bibliography{axial_collapse1_linearized_eqs}

\begin{thebibliography}{10}

\bibitem{Abrahams:1988vr}
A.~M. Abrahams and C.~R. Evans.
\newblock {Reading Off Gravitational Radiation Wave Forms in Numerical
  Relativity Calculations: Matching to Linearized Gravity}.
\newblock {\em Phys. Rev. D}, 37:318--332, 1988.

\bibitem{Abrahams:1992ib}
A.~M. Abrahams and C.~R. Evans.
\newblock {Trapping a geon: Black hole formation by an imploding gravitational
  wave}.
\newblock {\em Phys. Rev. D}, 46:R4117--R4121, 1992.

\bibitem{Abrahams:1993wa}
A.~M. Abrahams and C.~R. Evans.
\newblock {Critical behavior and scaling in vacuum axisymmetric gravitational
  collapse}.
\newblock {\em Phys. Rev. Lett.}, 70:2980--2983, 1993.

\bibitem{Abrahams:1994nw}
A.~M. Abrahams and C.~R. Evans.
\newblock {Universality in axisymmetric vacuum collapse}.
\newblock {\em Phys. Rev. D}, 49:3998--4003, 1994.

\bibitem{Baumgarte:2023tdh}
Thomas~W. Baumgarte, Bernd Br\"ugmann, Daniela Cors, Carsten Gundlach, David
  Hilditch, Anton Khirnov, Tom\'a\v{s} Ledvinka, Sarah Renkhoff, and
  Isabel~Su\'arez Fern\'andez.
\newblock {Critical Phenomena in the Collapse of Gravitational Waves}.
\newblock {\em Phys. Rev. Lett.}, 131(18):181401, 2023.

\bibitem{Baumgarte_Shapiro_2010}
Thomas~W. Baumgarte and Stuart~L. Shapiro.
\newblock {\em Numerical Relativity: Solving Einstein’s Equations on the
  Computer}.
\newblock Cambridge University Press, 2010.

\bibitem{Choptuik:1992jv}
Matthew~W. Choptuik.
\newblock {Universality and scaling in gravitational collapse of a massless
  scalar field}.
\newblock {\em Phys. Rev. Lett.}, 70:9--12, 1993.

\bibitem{Choptuik:2003as}
Matthew~W. Choptuik, Eric~W. Hirschmann, Steven~L. Liebling, and Frans
  Pretorius.
\newblock {An Axisymmetric gravitational collapse code}.
\newblock {\em Class. Quant. Grav.}, 20:1857--1878, 2003.

\bibitem{Evans:1984}
Charles~R. Evans.
\newblock {\em {A method for Numerical Relativity: simulation of axisymmetric
  gravitational collapse and gravitational radiation generation}}.
\newblock PhD thesis, The University of Texas at Austin, 1984.

\bibitem{Garfinkle:2000hd}
David Garfinkle and G.~Comer Duncan.
\newblock {Numerical evolution of Brill waves}.
\newblock {\em Phys. Rev. D}, 63:044011, 2001.

\bibitem{Hilditch:2017dnw}
David Hilditch, Andreas Weyhausen, and Bernd Br\"ugmann.
\newblock {Evolutions of centered Brill waves with a pseudospectral method}.
\newblock {\em Phys. Rev. D}, 96(10):104051, 2017.

\bibitem{Khirnov:2022rgw}
Anton Khirnov.
\newblock {\em {Representation of dynamical black hole spacetimes in numerical
  simulations}}.
\newblock PhD thesis, Charles U., Prague (main), 2022.

\bibitem{Ledvinka:2021rve}
Tom\'a\v{s} Ledvinka and Anton Khirnov.
\newblock {Universality of Curvature Invariants in Critical Vacuum
  Gravitational Collapse}.
\newblock {\em Phys. Rev. Lett.}, 127(1):011104, 2021.

\bibitem{Nollert:1999ji}
Hans-Peter Nollert.
\newblock {TOPICAL REVIEW: Quasinormal modes: the characteristic `sound' of
  black holes and neutron stars}.
\newblock {\em Class. Quant. Grav.}, 16:R159--R216, 1999.

\bibitem{Note1}
see Eqs. (22-28), (38-41) and (36,37) in \cite {Rostworowski:2017ruj}.

\bibitem{Note2}
first, from $tt$, $tr$, $tu$, $uu$ and $\varphi \varphi $ components of the
  metric we get Eqs. (\ref {eq:q1}-\ref {eq:q5}) and then, from the $rr$
  component, we get Eqs. (\ref {eq:hl2rr},\ref {eq:hl0rr}).

\bibitem{Note3}
After putting the Eq.(\ref {eq:xiltwou}) into a form \protect \[ \partial
  _r\left ( \protect \frac 1{r^3} \partial _r\protect \tilde {\xi }^{(\ell
  =2)}_u \right ) = \partial _r\left ( \protect \frac {\partial _r\Phi
  _2}{2r^2} + \protect \frac {\Phi _2}{r^3} \right ) \protect \,, \protect \]
  which is suitable for taking its 1st radial integral, we use the explicit
  form of $\Phi _2$: \protect \[ \Phi _2= f''_2(\protect \bar {u}) +
  g''_2(\protect \bar {v}) + 3 \protect \frac { f'_2(\protect \bar {u}) -
  g'_2(\protect \bar {v})}{r}+ 3 \protect \frac { f_2(\protect \bar {u}) +
  g_2(\protect \bar {v})}{r^2} \protect \, , \protect \] cf. Eq. (\ref
  {eq:masterScalar}).

\bibitem{Rinne:2008tk}
Oliver Rinne.
\newblock {Constrained evolution in axisymmetry and the gravitational collapse
  of prolate Brill waves}.
\newblock {\em Class. Quant. Grav.}, 25:135009, 2008.

\bibitem{Rostworowski:2017ruj}
Andrzej Rostworowski.
\newblock {Towards a theory of nonlinear gravitational waves: A systematic
  approach to nonlinear gravitational perturbations in the vacuum}.
\newblock {\em Phys. Rev. D}, 96(12):124026, 2017.

\bibitem{doi:10.1137/1.9780898719598}
Lloyd~N. Trefethen.
\newblock {\em Spectral Methods in MATLAB}.
\newblock Society for Industrial and Applied Mathematics, 2000.

\end{thebibliography}

\end{document}